%% file: sample631.tex
\documentclass[trackchanges,twocolumn]{aastex631}

\begin{document}

\title{Formation of the Two-Armed Phase Spiral from Multiple External Perturbations}

\author[0009-0002-4314-4138]{Junxian Lin}
\affiliation{Department of Astronomy, School of Physics and Astronomy, Shanghai Jiao Tong University, 800 Dongchuan Road, Shanghai, 200240, China}
\affiliation{State Key Laboratory of Dark Matter Physics, School of Physics and Astronomy, Shanghai Jiao Tong University, 800 Dongchuan Road, Shanghai, 200240, China}

\author[0000-0001-5017-7021]{Zhao-Yu Li}
\affiliation{Department of Astronomy, School of Physics and Astronomy, Shanghai Jiao Tong University, 800 Dongchuan Road, Shanghai, 200240, China}
\affiliation{State Key Laboratory of Dark Matter Physics, School of Physics and Astronomy, Shanghai Jiao Tong University, 800 Dongchuan Road, Shanghai, 200240, China}
\email{lizy.astro@sjtu.edu.cn}

\author[0009-0008-6155-3692]{Rui Guo}
\affiliation{National Astronomical Observatories, Chinese Academy of Sciences, Beijing 100101, China.}

\author[0000-0001-8917-1532]{Jason A. S. Hunt}
\affiliation{School of Mathematics $\&$ Physics, University of Surrey, Guildford GU2 7XH, UK}

\author[0000-0003-2595-5148]{T. Antoja}
\affiliation{Departament de Física Quàntica i Astrofísica (FQA), Universitat de Barcelona (UB), C Martí i Franquès, 1, 08028 Barcelona, Spain}
\affiliation{Institut de Ciències del Cosmos (ICCUB), Universitat de Barcelona (UB), C Martí i Franquès, 1, 08028 Barcelona, Spain}
\affiliation{Institut d’Estudis Espacials de Catalunya (IEEC), Edifici RDIT, Campus UPC, 08860 Castelldefels (Barcelona), Spain}

\author[0000-0001-6655-854X]{Chengye Cao}
\affiliation{Department of Astronomy, School of Physics and Astronomy, Shanghai Jiao Tong University, 800 Dongchuan Road, Shanghai, 200240, China}
\affiliation{State Key Laboratory of Dark Matter Physics, School of Physics and Astronomy, Shanghai Jiao Tong University, 800 Dongchuan Road, Shanghai, 200240, China}





\begin{abstract}
\input{abstract}
\end{abstract}


\section{Introduction} \label{sec:intro}
\input{introduction}

\section{Data and Method} \label{sec:data}
\input{data}

\section{Results} \label{sec:result}
\input{results}

\section{Discussion} \label{sec:discussion}
\input{discussion}

\section{Conclusion} \label{sec:conclusion}
\input{conclusion}


\section{Acknowledgments} \label{sec:ack}
This work is supported by the National Natural Science Foundation of China under grant No. 12233001, by the National Key R$\&$D Program of China under grant No. 2024YFA1611602, by a Shanghai Natural Science Research Grant (24ZR1491200), by the ``111'' project of the Ministry of Education under grant No. B20019, and by the China Manned Space Project with No. CMS-CSST-2025-A11. We thank the sponsorship from Yangyang Development Fund. JH is supported by an STFC Ernest Rutherford Fellowship (ST/Z510245/1). TA acknowledges the grants PID2021-125451NA-I00 and CNS2022-135232 funded by MICIU/AEI/10.13039/501100011033 and by ``ERDF A way of making Europe’’, by the ``European Union'' and by the ``European Union Next Generation EU/PRTR'', and the Institute of Cosmos Sciences University of Barcelona (ICCUB, Unidad de Excelencia ’Mar\'{\i}a de Maeztu’) through grant CEX2019-000918-M. RG is supported by Initiative Postdocs Supporting Program (No. BX2021183), funded by China Postdoctoral Science Foundation, and by the National Natural Science Foundation of China under grant No. 12103031. This work made use of the Gravity Supercomputer at the Department of Astronomy, Shanghai Jiao Tong University. 

This work has made use of data from the European Space Agency(ESA) mission Gaia (\url{https://www.cosmos.esa.int/gaia}), processed by the Gaia Data Processing and Analysis Consortium (DPAC, \url{https://www.cosmos.esa.int/web/gaia/dpac/consortium}). Funding for the DPAC has been provided by national institutions, in particular the institutions participating in the Gaia Multilateral Agreement. 

\bibliography{sample631}{}
\bibliographystyle{aasjournal}



\end{document}

%% file: abstract.tex
Recent studies using the Gaia DR3 data have revealed a two-armed phase spiral in the $Z-V_Z$ phase space in the inner disk. In this study, we present new features of the two-armed phase spiral revealed by the Gaia Data and a new mechanism to explain such features with multiple external perturbations. By segmenting the Gaia DR3 RVS catalog based on $J_{\phi}$ (or $R_{g}$) and $\theta_\phi$, we confirm the existence of the clear two-armed phase spiral in the inner disk. Moreover, we identify a different two-armed phase spiral pattern at slightly larger radii, resembling a weak secondary branch along with the prominent major branch. At a given radius, with the azimuthal angle increasing, we observe a systematic transition of the two-armed phase spiral, with the significance of one branch weakened and another branch enhanced. This two-armed phase spiral may be due to the overlapping of distinct one-armed phase spirals. At different radii, the perturbation times estimated from each branch of the two-armed phase spiral are $\sim 320$ Myr and $\sim 500$ Myr, respectively, suggesting that the Galactic disk could be impacted by double external perturbers separated by $\sim 180$ Myr. We also performed test particle simulations of the disk perturbed by two satellite galaxies, which successfully generated a two-armed phase spiral similar to the observation.  Both the observation and simulation results suggest that the signature in the $Z-V_Z$ phase space of earlier perturbations may not be completely erased by the more recent one. 

%% file: introduction.tex
The phase space structures in the Milky Way galaxy reveal important physical processes of the Galactic structure formation and evolution. Evidence of the Milky Way disk disequilibrium arises from vertical density asymmetries, high-order kinematic modes (i.e., bending and breathing modes), flares, warps of the Galactic disk as well as large-scale in-plane spiral structure \citep[][]{Widrow2012, Williams2013, Carlin2013, Xu2015, Sun2015, Chen2019, Poggio2020, Poggio2021, Huang2024}. The Gaia satellite has revolutionized our understanding of the Milky Way by providing high-precision stellar position and velocity measurements for billions of stars. One important result in the Gaia Data Release 2 (DR2) \citep{Gaia2018} is the discovery of complex structures within multiple phase spaces, i.e., arches in the $V_R-V_\phi$ space, ridge lines in the $R-V_\phi$ space, and spirals in the $Z-V_Z$ space \citep{Antoja2018}. In particular, the existence of the phase spiral in the $Z-V_Z$ space (also called phase snail shell) suggests that the Milky Way recently experienced an external perturbation \citep[][]{Antoja2018, Laporte2018, Binney2018, Bland2021}. The Sagittarius dwarf (Sgr) galaxy is widely believed to be the main driver of the phase spiral \citep{Antoja2018, Laporte2019}, although the exact origin is still under debate\citep{Bennett2022, Tremaine2023, GarciaConde2024,Hunt2025}.

The phase spiral has emerged as an important feature for understanding the merger history and intrinsic properties of the Milky Way. 
Previous works have suggested that the perturbation happened $\sim$ 300 to 1000 Myr ago \citep{Antoja2018,Tian2018,Binney2018,Bland2019,LiZ2020,LiZ2021,LiH2021,Elise2023,Antoja2023,Frankel2023}. Connections between N/S asymmetries and the phase spiral have also been suggested \citep{Gomez2013,Salomon2020, Guo2022, LinJ2024}. Moreover, the phase spiral can be used to infer the Galactic disk potential \citep{Widmark2021a, Widmark2021b,Widmark2022b, Widmark2022a}. In particular, without assuming the overall analytical functional form of the Galactic potential, \citet{Guo2024} simply used the geometric shape of the phase spiral to directly measure the vertical potential profile at the different radii of the Galactic disk. They further analyzed the vertical potential profile in the solar neighborhood to indicate a local dark matter density of $\rho_{dm} = 0.0150 \pm 0.0031 \text{M}_{\odot}\text{pc}^{-3}$, which is consistent with the previous results \citep{LiH2021, LiH2023,deSalas2021, Guo2020}. 



With the Gaia DR3 data, \citet{Hunt2022} first showed the existence of a two-armed phase spiral pattern in the inner disc by binning the data according to the guiding center radius ($R_{g}$) and the conjugate angle to the azimuthal action ($\theta_{\phi}$). Their simulations suggested that the formation of the two-armed phase spiral is likely related to the bar or spiral structure. \citet{LiC2023} used the test particle simulation of a decelerating bar to generate the two-armed phase spiral, first appearing in the inner disc, and later propagating to larger radii. These studies indicate that the internal perturbation from the bar may play a significant role in the two-armed phase spiral formation. In addition,  \citet{Widrow2023} demonstrated theoretically that a cloud on a circular orbit within the disc could also excite the two-armed phase spiral pattern. Recently, \citet{Asano2025} showed that the two-armed phase spiral can be generated by the breathing mode of the spiral arms, which is triggered by the single satellite perturbation in $N$-body simulation. \citet{Chiba2025} also demonstrated that a two-armed phase spiral can be formed and maintained in steady rotation for the resonance, driven by both periodic and stochastic perturbations (e.g., spiral arms and giant molecular clouds).


However, there are still important questions unanswered regarding to the two-armed phase spiral. For example, how does the two-armed phase spiral form and evolve? Does the two-armed phase spiral exist in the other part of the disk? Apparently, more efforts are needed from both the observational and theoretical perspectives. In this paper, we first re-examine the two-armed phase spiral properties thoroughly in the observation. Then we utilize a test particle simulation to investigate a new scenario, whether the two-armed phase spirals could be generated by multiple external perturbations. In this work, we test the hypothesis that the observed two-armed phase spiral may result from a composite of two distinct one-armed phase spirals due to multiple external perturbations, rather than an intrinsically symmetric structure resulting from a breathing mode.

The paper is organized as follows. Section \ref{sec:data} shows the data selection and the method to measure the phase spiral shape. In Section \ref{sec:result} we examine the observational results of the two-armed phase spiral in detail. The test particle simulation results are shown in Section \ref{sec:sim}. These results are discussed in Section \ref{sec:discussion} and summarized in Section \ref{sec:conclusion}.


%% file: data.tex
For the observation, we select the Gaia DR3 RVS sample with a relative parallax error of less than 20$\%$. The Bayesian distance from \citet{Bailer-Jones2021} is used. We use \textit{astropy} \citep{astropy2022} to transform the stellar celestial coordinates into the Galactocentric Cartesian and cylindrical coordinates.  The Galactocentric radius of the Sun is taken to be $R_{\odot}=8.275$ kpc \citep{Gravity2021}, with the height of the Sun above the Galactic midplane $z_{\odot}=0.0208$ kpc \citep{Bennett2019}. Additionally, we adopt the peculiar velocity of the Sun with respect to the Local Standard of Rest (LSR) as $(u_{\odot},v_{\odot},w_{\odot})=(11.1,12.24,7.25)$ km s$^{-1}$ \citep{Schonrich2010} with the LSR velocity $v_{\text{LSR}}=238$ km s$^{-1}$ \citep{Schonrich2012}.













For the stars in our sample, we then compute the actions ($J_{r},J_{\phi},J_{z}$), the conjugate angles ($\theta_{r},\theta_{\phi},\theta_{z}$) and the frequencies ($\Omega_{r},\Omega_{\phi},\Omega_{z}$)  in the cylindrical coordinate by \textit{AGAMA} \citep{Vasiliev2019}, with the MWPotential2014 potential model \citep{Bovy2015}. Based on the rotation curve derived from MWPotential2014, we can estimate the guiding center radius: $R_g \simeq J_\phi/v_c$. Following \citet{Hunt2020}, we also define the cartesian coordinates in the action space: $x_{act} = R_{g} \text{cos}\theta_\phi$, $y_{act} = R_g \text{sin}\theta_\phi$.

\begin{figure}[ht!]
    \plotone{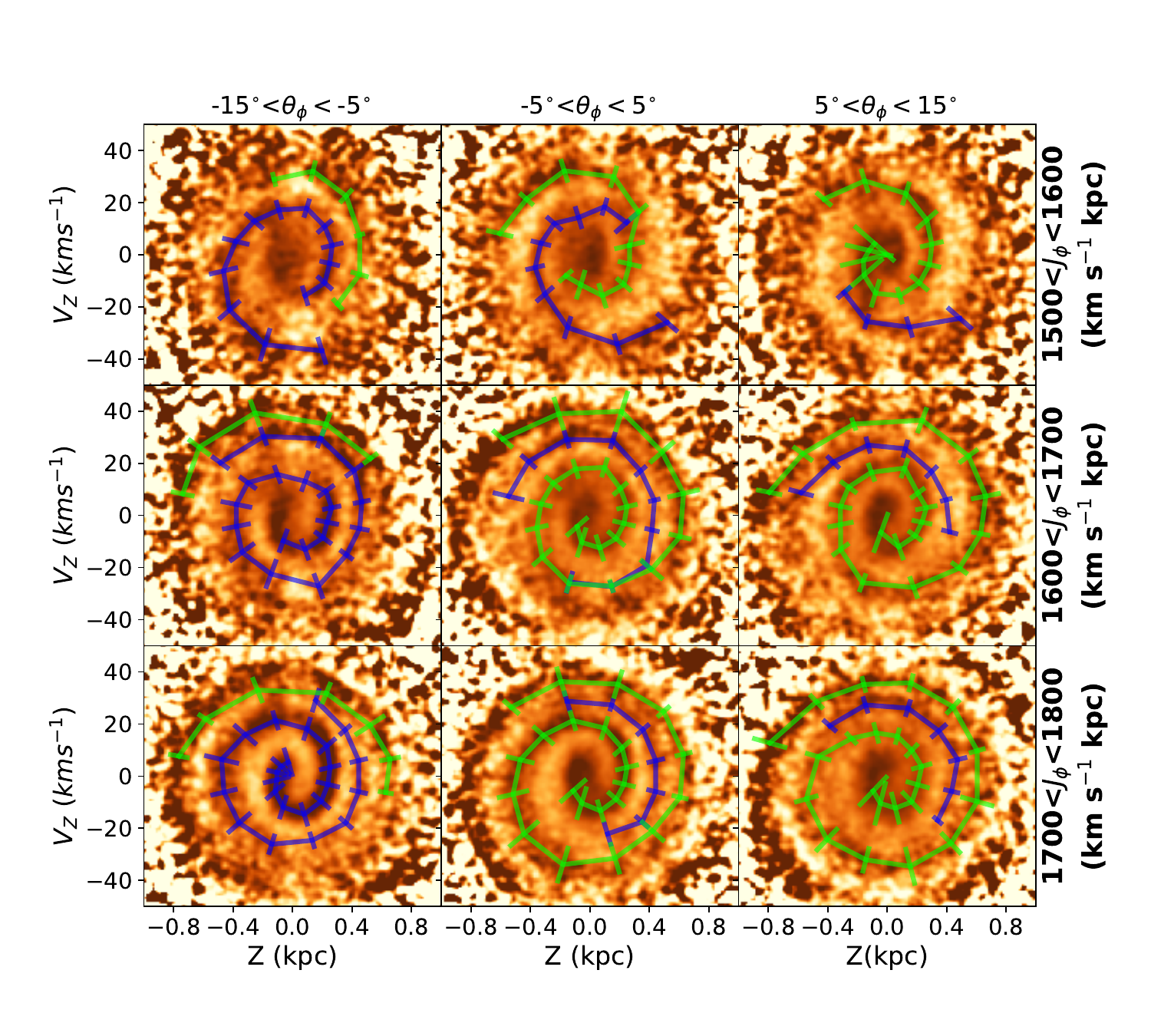} 
    \setlength{\abovecaptionskip}{-0.07\linewidth}
    \flushleft

    \caption{Density contrast map $\Delta N$ of the $Z-V_Z$ phase space with $J_\phi$ increasing from top to bottom rows and $\theta_\phi$ increasing from left to right columns. In each panel, the blue and green lines represent the major and secondary branches. Note that the major (or secondary) branch in the left column becomes the secondary (or major) branch in the right column.}
    \label{fig:obs_1}
\end{figure}

To enhance the phase spiral feature in the $Z-V_{z}$ space, similar to previous works \citep[e.g.,][]{Laporte2019, LiZ2020}, we generate the number density contrast map ($\Delta N$) for the phase space\footnote{$\Delta N=(N-N_b)/N_b$, where the background $N_b$ is the Gaussian kernel convolved number density map of the phase space.}. In the following analysis, we measure the shape of the phase spiral in the number density contrast map using methods similar to \citet{LiZ2020} and \citet{LiZ2021}. Briefly speaking, the $Z-V_Z$ phase space is azimuthally divided into 12 equally spaced wedges with the origin at the center of the phase space ($Z$, $V_Z$) = (0,0). Within each wedge, along the radial direction in the $Z-V_Z$ phase space, the peak positions of the $\Delta N$ map are identified and connected sequentially along the azimuthal direction to trace the geometric shape of the phase spiral, with half the separation between the two local minimum positions in the radial direction of the phase space adjacent to each peak as the width of the phase spiral. In practice, the measured peak positions in the outermost region of the $Z-V_Z$ phase space are discarded because of the much lower stellar number density resulting in the $\Delta N$ map.

%% file: results.tex


\subsection{Two-Armed Phase Spiral in Gaia DR3}

Similar to \citet{Hunt2022}, the stars are binned into different $J_\phi$ and $\theta_\phi$ ranges to identify the phase space structures. We have found two-armed phase spirals with $\text{1500}<J_\phi<\text{1600}$ km s$^{-1}$ kpc as shown in Fig.\,\ref{fig:obs_1}. The top row reveals the phase spiral in the inner disk at $R_g \approx 6$ kpc ($\text{1500}<J_\phi<\text{1600}$ km s$^{-1}$ kpc) within $-15^{\circ}<\theta_\phi<15^{\circ}$, consistent with \citet{Hunt2022}. Specifically, the middle panel with $-5^{\circ}<\theta_\phi<5^{\circ}$ shows the most prominent two-armed phase spiral compared to the other azimuthal ranges. From top to bottom rows as $J_\phi$ increases, the two-armed phase spiral transitions to a single-arm spiral with a weaker secondary branch, which has also been reported in \citet{Antoja2023}. 

\begin{figure}[ht!]
    \hspace{-1.2cm}\includegraphics[width=0.6\textwidth,height=0.2\textwidth]{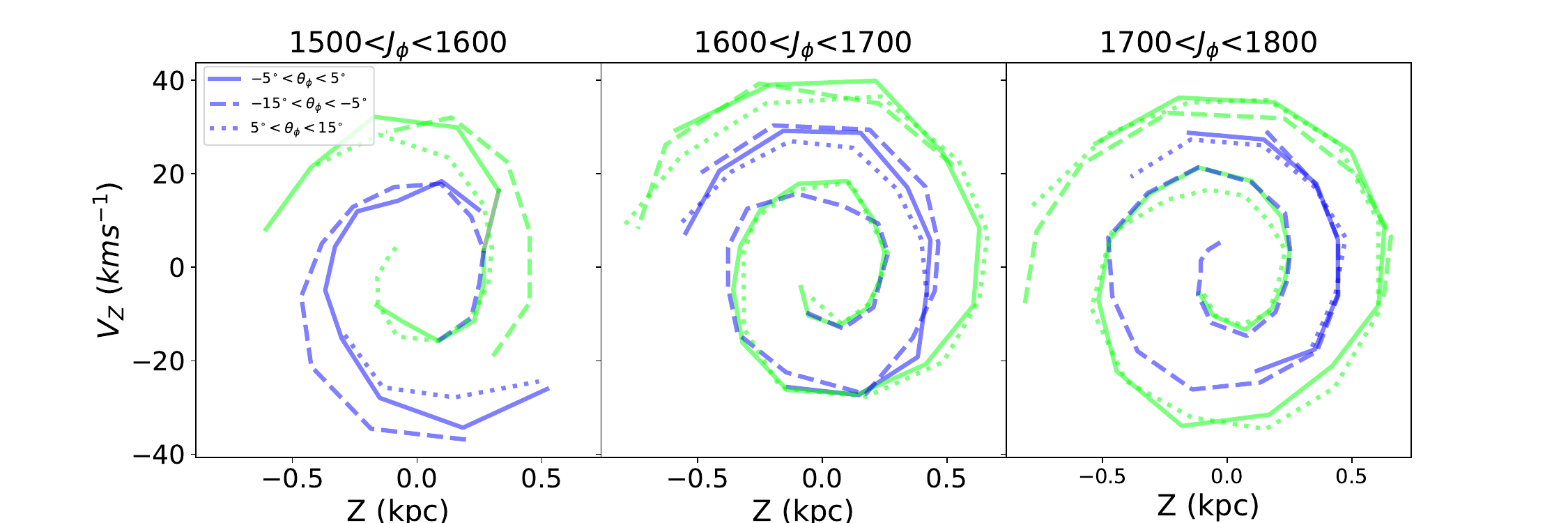} 

    \caption{Comparing the measured geometric shapes of the two-armed phase spirals in different $J_\phi$ ranges. In each panel, the blue (or green) curves represents the branch of the phase spiral with the same color as shown in Fig.\,\ref{fig:obs_1}, with different types of lines for different azimuthal ranges.}
    \label{fig:obs_2}
\end{figure}

To better compare the phase spirals in different azimuthal ranges, we measure the geometric shapes of the phase spirals in Fig.\,\ref{fig:obs_1}, with the blue and green curves representing the two branches. Comparing the phase spirals with the same $J_\phi$ value but different $\theta_\phi$ values (from left to right columns), we find a systematic variation of the significance of each branch. The major (or secondary) phase spiral branch in the left column with $-15^{\circ}<\theta_\phi<-5^{\circ}$ becomes the secondary (or major) phase spiral branch in the right column with $5^{\circ}<\theta_\phi<15^{\circ}$. In this measurement, different colors correspond to different major phase spiral branches in each azimuthal range. To avoid confusion, we refer to the major phase spiral branch in $-15^{\circ}<\theta_\phi<-5^{\circ}$ and the secondary phase spiral branch in $5^{\circ}<\theta_\phi<15^{\circ}$ as the blue branch, as both are represented by the blue curve in Fig.\,\ref{fig:obs_1}. Similarly, the green branch corresonds to the phase spiral branch traced by the green curve.

\begin{figure}[htbp]
    \plotone{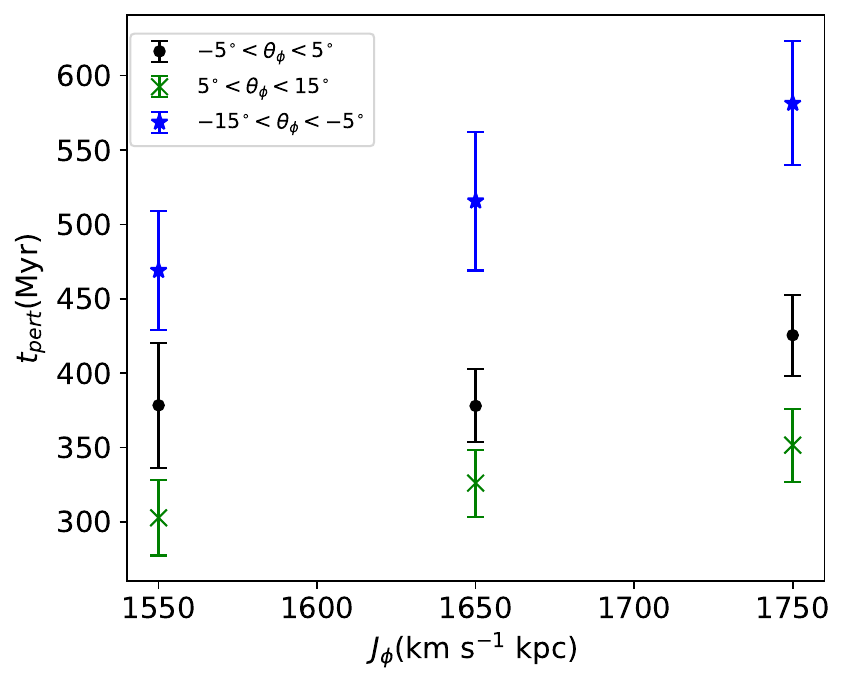}
    \caption{The perturbation times derived from the two-armed phase spirals in different azimuthal ranges shown in Fig.\,\ref{fig:obs_1} in different azimuthal ranges. The blue points correspond to the blue branches of phase spirals in $-15^{\circ}<\theta_\phi<-5^{\circ}$ (the major branch in the left column of Fig.\,\ref{fig:obs_1}), and the green points correspond to the green branches of phase spirals in $5^{\circ}<\theta_\phi<15^{\circ}$ (the major branch in the right column of Fig.\,\ref{fig:obs_1}). The black dots represent the perturbation times derived by fitting both the blue and green branches of the phase spirals in $-5^{\circ}<\theta_\phi<5^{\circ}$ simultaneously.}
    \label{fig:est_time}
\end{figure}

Based on our observational results, it seems that such an  `apparent' two-armed phase spiral could be a composite of two different sets of single-arm phase spirals. Note that this is different from the `intrinsic' (symmetic) two-armed phase spirals generated by the bar-induced breathing mode. In Fig.\,\ref{fig:obs_2} we directly compare the shapes of the two-armed phase spirals in different azimuthal ranges. In each panel, the blue (or green) branches from different azimuthal ranges have similar shapes, with a slight mismatch reflecting the azimuthal variation of the phase spiral \citep{Alinder2023,Antoja2023}. In the central region of the phase space with $|Z| < 0.2$ kpc and $|V_Z| < 20$ km s$^{-1}$, the blue and green branches merge together, with the primary difference in the outer part of the phase space.

\subsection{Perturbation Time Estimation from Phase Spirals}

Theoretically speaking, the phase spiral in $Z-V_Z$ space corresponds to a series of approximately parallel lines in the $\theta_{z}-\Omega_{z}$ space, where $\Omega_z$ and $\theta_z$ are the vertical oscillation frequency and phase angle, respectively. The slope of the $\theta_z - \Omega_z$ line represents the perturbation time ($t_{\rm pert}$). Similar to the previous works \citep{LiH2021,Elise2023, Guo2024, Widmark2021a, Widmark2021b, Widmark2022b, Widmark2022a}, we estimate the perturbation time by fitting the equation

\begin{equation}
    \theta_{z} = \Omega_{z} t_{\rm pert} + \theta_{z,0}, 
    \label{eq:est_time}
\end{equation}
where $t_{\rm pert}$ is the perturbation time, and $\theta_{z,0}$ is the initial phase angle. 
The shape of the measured $Z-V_Z$ phase spiral in Fig.\,\ref{fig:obs_1} is converted into the $\theta_z-\Omega_z$ space assuming the Galactic potential MWPotential2014 \citep{Bovy2015}. Then the perturbation time is estimated by fitting Equation \ref{eq:est_time} to the $\theta_z-\Omega_z$ line.

Fig.\,\ref{fig:est_time} shows the perturbation times measured from the phase spirals in Fig.\,\ref{fig:obs_1}. The blue points correspond to the results from the blue branch of the phase spiral in the $-15^{\circ}<\theta_\phi<-5^{\circ}$ region (i.e., the major branch in the left column of Fig.\,\ref{fig:obs_1}), while the green points are derived from the green branch of the phase spiral in the $5^{\circ}<\theta_\phi<15^{\circ}$ region (i.e., the major branch in the right column). The perturbation times derived by simultaneously both branches in $-5^{\circ}<\theta_\phi<5^{\circ}$ are shown with the black dots. The blue points are systematically higher than the green points, with the black points in between. This suggests that the two branches correspond to different perturbation times. Regardless of the $J_\phi$ ranges in this work, the blue branches of the phase spirals in the left column of Fig.\,\ref{fig:obs_1} (i.e.,  $-15^{\circ}<\theta_\phi<-5^{\circ}$) have a similar perturbation time of $\sim$ 500 Myr (blue points), while the green branches of the phase spirals in the right column (i.e.,  $5^{\circ}<\theta_\phi<15^{\circ}$) have $t_{pert} \sim 320$ Myr (green points), with the black dots of $t_{pert} \sim 380$ Myr. It has been suggested that when neglecting self-gravity, the perturbation time tends to be underestimated \citep{Widrow2023}. Therefore, rather than constraining the exact perturbation time, our results highlight the differences in perturbation times of the two different branches across the azimuthal ranges.

The systematic difference in the perturbation time between the green and blue points indicates that the Milky Way may have recently experienced two consecutive external perturbation events separated by $\sim$ 180 Myr. 
The perturbation time derived from the blue points ($\sim$ 500 Myr) is consistent with the previous works based on the single arm phase spirals at the outer radii \citep[e.g., ][]{LiZ2021, Antoja2023}. On the other hand, the green points ($\sim$ 320 Myr) suggest a more recent perturbation event. Similarly, \citet{Elise2023} measured the shape of the two-armed phase spiral within $\text{1500}<J_\phi<\text{1600}$ km s$^{-1}$ kpc and even find a more recent perturbation with $t_{\rm pert} \sim 133$ Myr. It is possible that a more recent perturbation may not be strong enough to fully erase the phase space signature in the inner disk left by the previous perturbation, contributing to the formation of the two branches in the observed phase spiral.

\section{Test Particle Simulation}  \label{sec:sim}

\begin{figure}[ht!]
    \centering
    \plotone{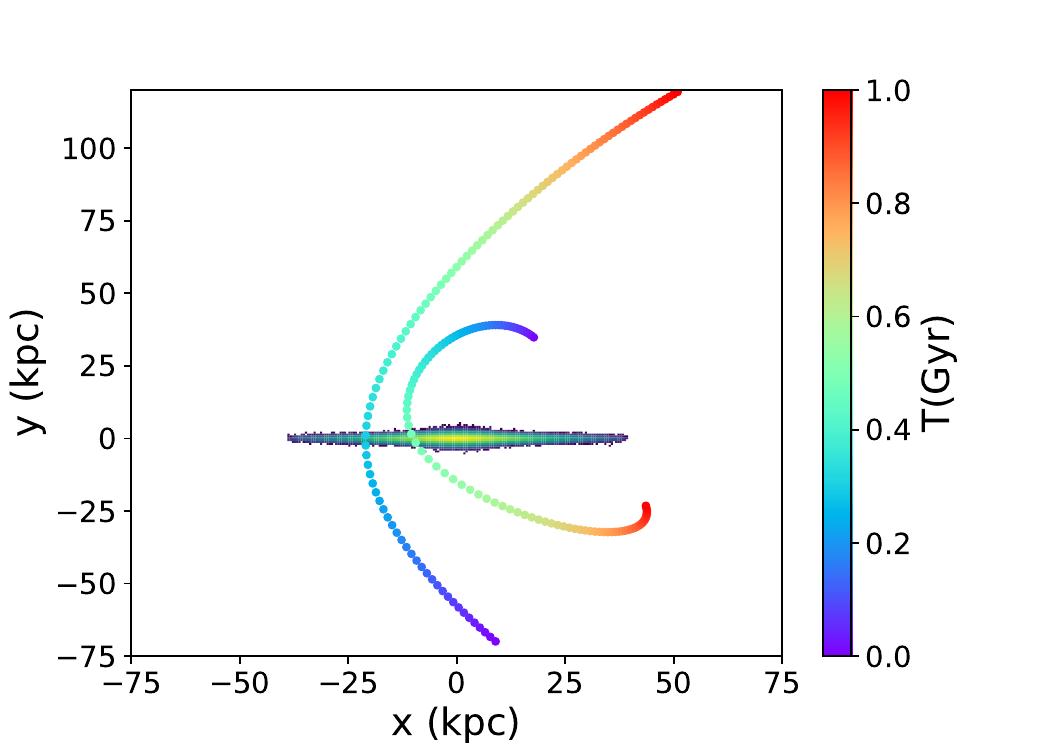}
    \caption{The setup of the multiple perturbation scenario in the test particle simulation, where the color represents the simulation time. The first satellite with a mass of $2.5 \times 10^{10}$ $M_{\odot}$ will encounter the Milky Way at $(x,y,z) = (-21,-4,0)$kpc at 300 Myr. After a further 180 Myr, the second perturbation occurs at $(x,y,z) = (-10,4,0)$ kpc with the satellite mass of $1 \times 10^{10}$ $M_{\odot}$.}
    \label{fig:orb}
\end{figure}

\begin{figure*}[ht!]
    \centering
    \plotone{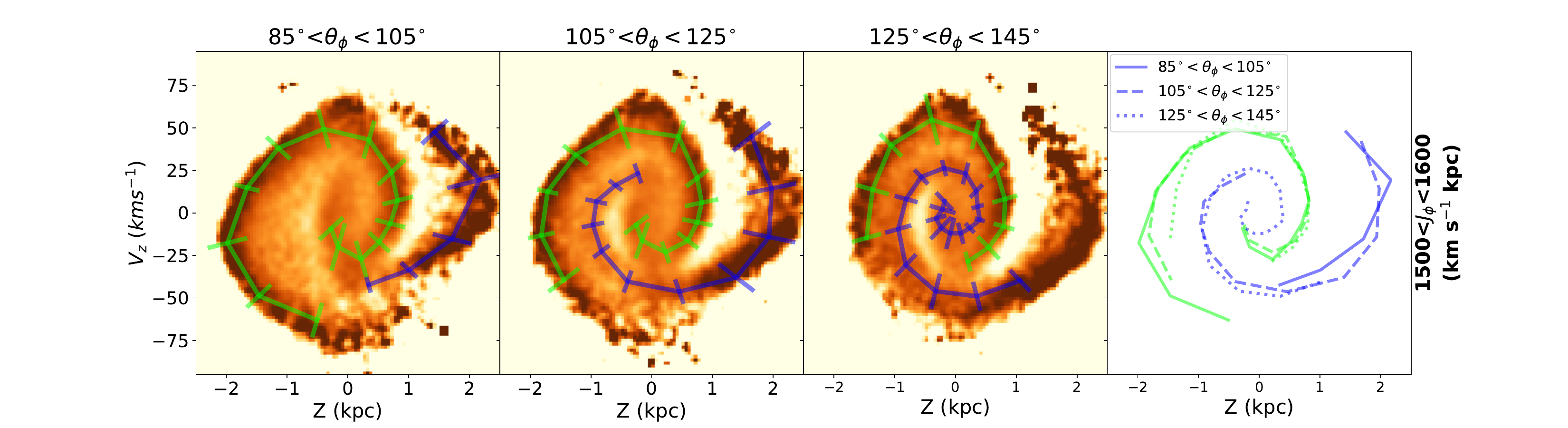}
    \caption{The $Z-V_Z$ phase space number density maps for stars with $1500<J_{\phi}<1600$ of the test particle simulation at 200 Myr after the secondary passage of the satellite perturbation at $85^{\circ}<\theta_\phi<105^{\circ}$, $105^{\circ}<\theta_\phi<125^{\circ}$, and $125^{\circ}<\theta_\phi<145^{\circ}$ shown in the left three panels. The blue and green curves represent the measured branches of the phase spirals in the left three panels. These curves are overlaid in the right panel for comparison. The blue (green) branches in different azimuthal angles are generally consistent.}
    \label{fig:sim_re}
\end{figure*}

\begin{figure}[ht!]
    \centering
    \plotone{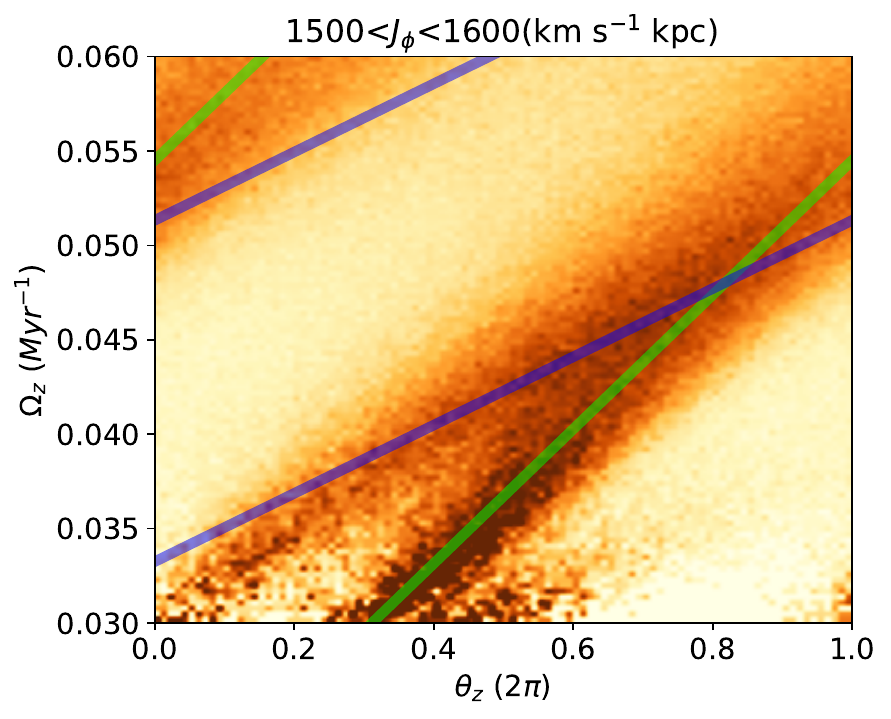}
    \caption{The $\theta_z-\Omega_z$ space corresponding to the $Z-V_Z$ phase space of the test particle simulation with $1500<J_{\phi}<1600\, \text{km}\, \text{s}^{-1}\, \text{kpc}$ (without separating $\theta_\phi$) at 200 Myr after the second passage of the satellite. The green and blue lines are the best fit linear relations for the two branches of the phase spirals, denoted by the green and blue curves in Fig.\,\ref{fig:sim_re}, respectively. The slope of the best-fitting linear relationship is the inverse of the perturbation time ($1/t_{pert}$). The estimated perturbation times of the green and blue curve are $\sim 200$ Myr and $\sim 350$ Myr, respectively}
    \label{fig:sim_OT}
\end{figure}
In this section, we use the test particle simulation to explore the possibility of the formation of the apparent two-armed phase spiral with multiple external perturbations. We use \textit{AGAMA} \citep{Vasiliev2019} to construct the initial distribution function and perform the orbit integration in the Milky Way potential MWPotential2014. We adopt the  QuasiIsothermal distribution function \citep{Binney2010,Binney2011} to represent a cold thin disc with the radial scale length of 3.7 kpc, and the scale height of 0.3 kpc, using 10 million particles \citep{Binney2015,Bland2016}. Although we neglect self-gravity, the simulation of test particles can still reveal important physical mechanisms underlying the observed phenomenon \citep[e.g.,][]{Antoja2018, LiZ2021, Antoja2022, LiC2023, Cao2024}.


We test the aforementioned scenario by introducing two consecutive external satellite perturbations. A simplified perturbation setup is adopted, where each satellite is treated as a point mass moving through the fixed Milky Way potential, with the trajectories shown in Fig.\,\ref{fig:orb}. The pericenter of the first satellite is at $(x,y,z) = (-21,-4,0)$ kpc with a velocity of $(V_x, V_y, V_z) = (4,-111,337)$ km s$^{-1}$ \citep{Vega2015}, and a mass of $2.5 \times 10^{10}$ $M_{\odot}$. After approximately 180 Myr, similar to the perturbation time difference measured from different branches shown in  Fig.\,\ref{fig:est_time}, a second satellite of mass $1 \times 10^{10} M_{\odot}$ crosses the disk at $(x,y,z) = (-10,4,0)$ kpc with velocity $(V_x, V_y, V_z) = (100, 155, -300)$ km s$^{-1}$. The orbit and mass of the second satellite are artificially designed to create an apparent two-armed phase spiral feature without completely erasing the signature of the previous perturbation. This setup is not intended to represent any particular satellite of the Milky Way, but rather to explore how consecutive external perturbations may superpose to generate complex vertical phase space structures.

This test particle simulation with the specific setup allows us to investigate how the overlapping kinematic responses from multiple perturbations can lead to an apparent two-armed phase spiral, , qualitatively similar to the Gaia data. The chosen time separation of $\sim$180 Myr between two consecutive perturbations qualitatively matches the time difference between the two branches of the phase spirals in different azimuthal sectors (see Fig.\,\ref{fig:est_time}). Our approach is similar to the previous analytical and test-particle studies \citep[e.g.,][]{DOnghia2010}, where satellite-induced vertical impulses generated coherent structures in the phase space even in the absence of self-gravity. In this context, the resulting phase spiral is not a self-gravitating structure, but a coherent response retained in the phase space.

 \begin{figure*}[ht!]
    \centering
    \plotone{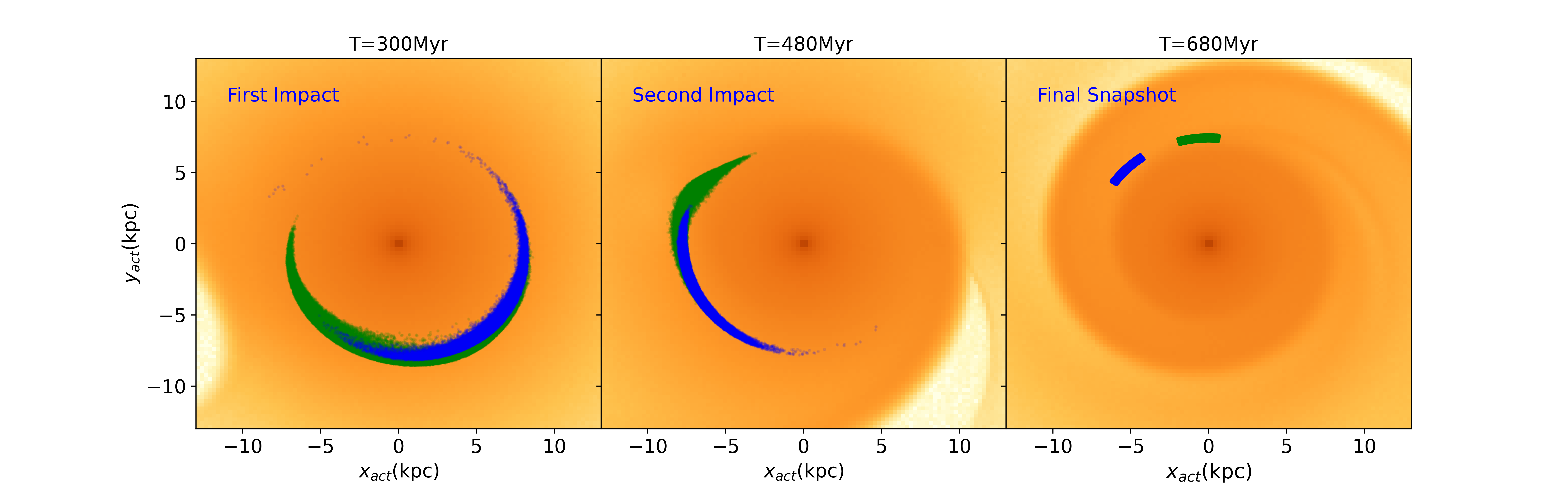}
    
    \caption{The face-on view ($x_{\text{act}}-y_{\text{act}}$) of the simulation at the time of the first impact (left panel: T = 300Myr), the second impact (second panel: T = 480Myr), and the final snapshot (right panel: T = 680 Myr). The green and blue dots represent the test particles in the final snapshot with $\text{1500}<J_\phi<\text{1600}$ and $85^{\circ}<\theta_\phi<105^{\circ}$ and $125^{\circ}<\theta_\phi<145^{\circ}$, respectively. The left and middle panels trace the green and blue particles back to the times of the first and second perturbations.}
    \label{fig:6}
\end{figure*}

\begin{figure*}[ht!]
    \centering
    \plotone{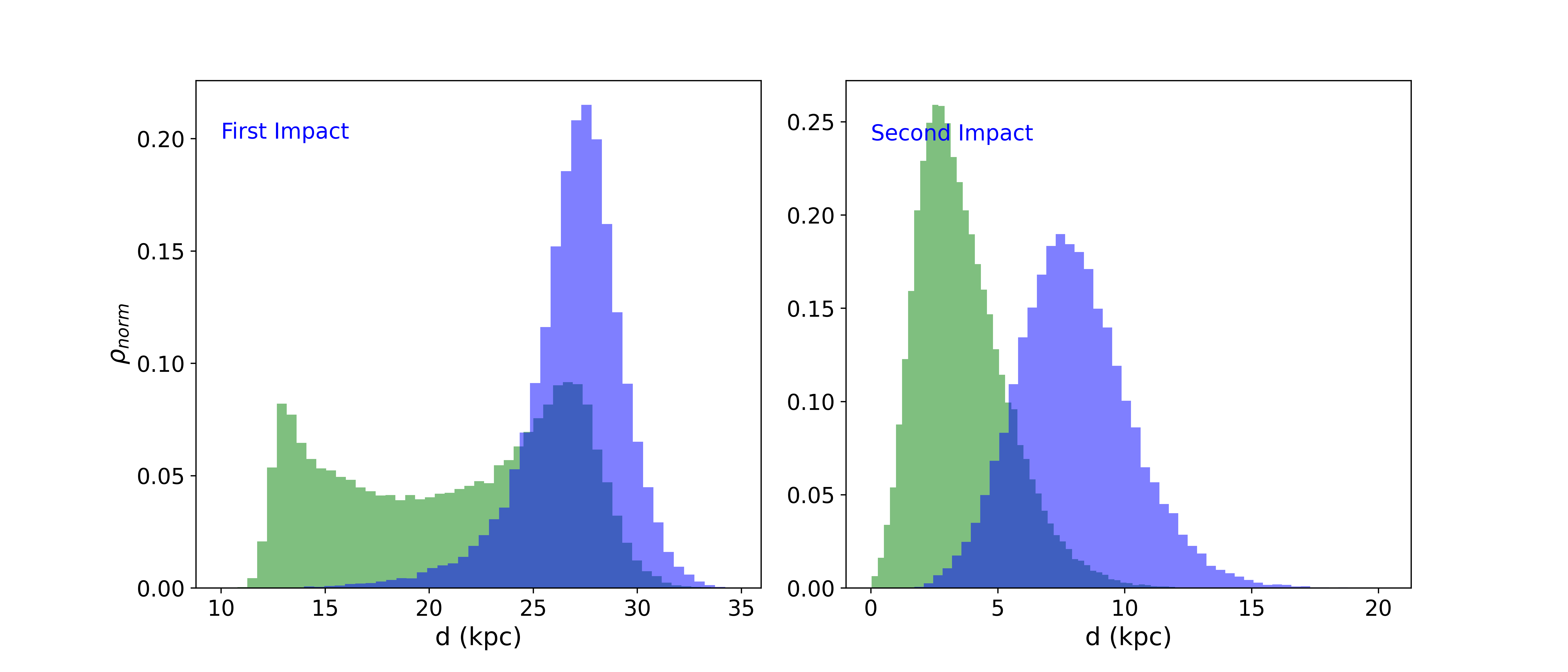}
    \setlength{\abovecaptionskip}{-0.01\linewidth}
    \caption{The distance distributions of green and blue particles relative to the impact sites of the first (left) and second (right) impacts. In the second impact, the green particles are located much closer to the impact site compared to the blue particles.}
    \label{fig:7}
\end{figure*}

Fig.\,\ref{fig:sim_re} shows examples of the apparent two-armed phase spirals in three adjacent $\theta_{\phi}$ ranges of the inner disk ($1500<J_\phi <1600\, \text{km}\, \text{s}^{-1}\, \text{kpc}$) at $200$ Myr after the impact of the second satellite. In each panel, we identify and measure both branches of the phase spirals, represented by the green and blue curves, respectively. Notably, the major branch of the phase spiral in the third panel ($125^{\circ}<\theta_\phi<145^{\circ}$) remains discernible as a weak secondary branch in the first panel ($85^{\circ}<\theta_\phi<105^{\circ}$). A similar branch is also observed in the second panel ($105^{\circ}<\theta_\phi<125^{\circ}$), contributing to the apparent two-armed phase spiral. Similarly, when we overlay the measured branches of these phase spirals together in the fourth panel, we find good agreement between different azimuthal angles, qualitatively similar to the observational results in Figs.\,\ref{fig:obs_1}, and \ref{fig:obs_2}. This test supports the hypothesis that the apparent two-armed phase spiral could originate from multiple external perturbations. Such an apparent two-armed phase spiral is generated by overlapping of the multiple bending modes, which is different from the intrinsic two-armed phase spirals arising from the breathing mode generated by the bar.

 \begin{figure*}[ht!]
    \centering
    \plotone{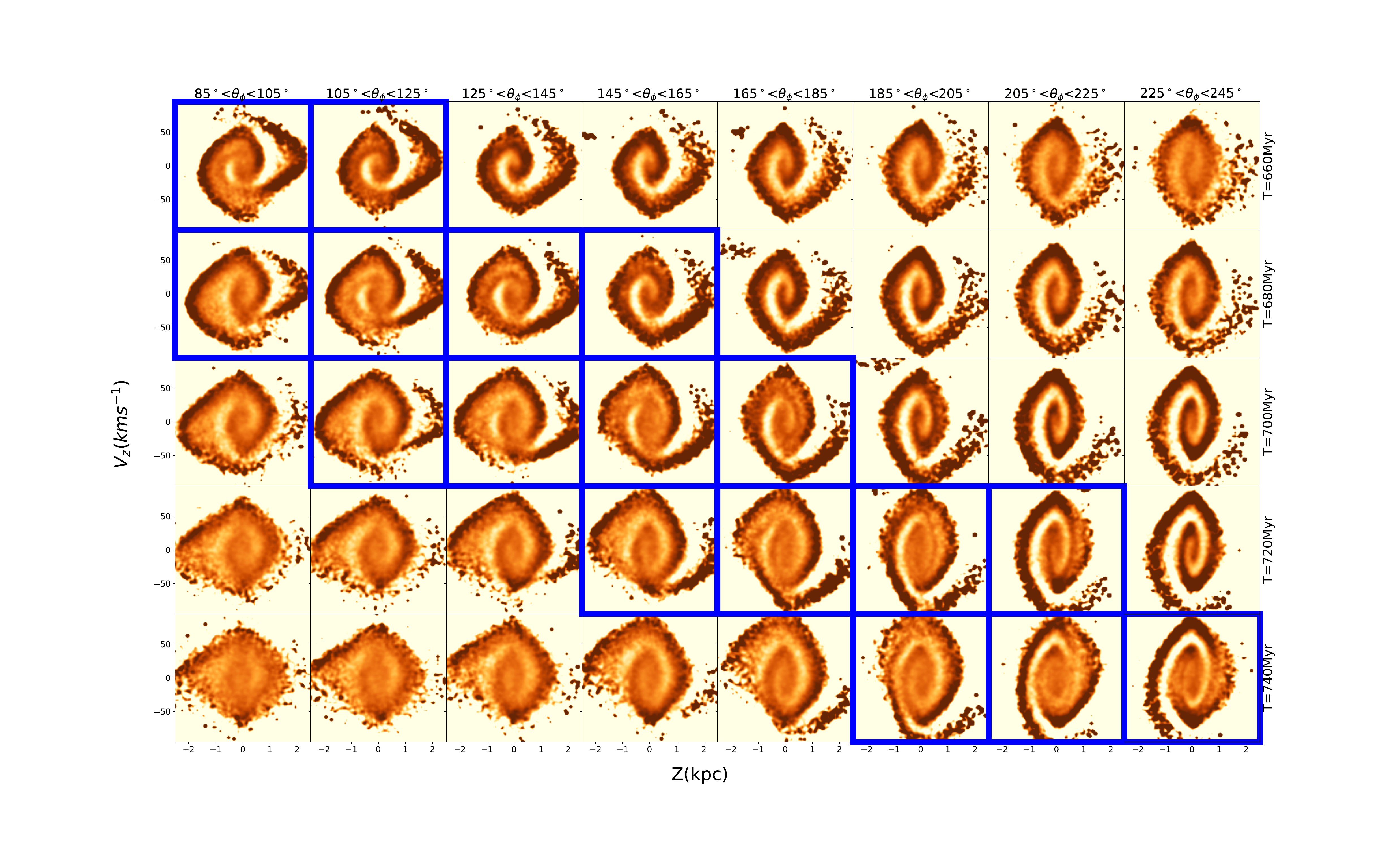}
    \setlength{\abovecaptionskip}{-0.07\linewidth}
    \caption{Evolution of the apparent two-armed phase spirals in different azimuthal ranges with $1500<J_\phi <1600\, \text{km}\, \text{s}^{-1}\, \text{kpc}$. Panels with the apparent two-armed phase spiral (or secondary branch) are highlighted with blue boxes. From top to bottom rows, the time increases from 660 Myr to 740 Myr. The apparent two-armed phase spiral becomes less clear at later times, resulting in a single-armed phase spiral with a weaker secondary branch.}
    \label{fig:8}
\end{figure*}

By transforming these phase spirals in Fig.\,\ref{fig:sim_re} into the $\theta_z - \Omega_z$ space in Fig.\,\ref{fig:sim_OT} (without separating different $\theta_\phi$ ranges), we try to estimate the perturbation time from each branch of the phase spiral. As shown in Fig.\,\ref{fig:sim_OT}, the stripes with different slopes in the $\theta_z - \Omega_z$ space correspond to the two different branches of the phase spirals in Fig.\,\ref{fig:sim_re}. The blue (or green) line in the $\theta_z-\Omega_z$ space is the best fit linear relation of the corresponding blue (or green) phase spiral. For the blue line ($125^{\circ}<\theta_\phi<145^{\circ}$), the slope of the linear relation in the $\theta_z-\Omega_z$ space indicates the perturbation time of $\sim$ 350 Myr, while the perturbation time inferred from the green line ($85^{\circ}<\theta_\phi<105^{\circ}$) is $\sim$ 200 Myr. These values exhibit a time difference of 150 Myr, roughly consistent with the configuration of the simulation with two consecutive external perturbations separated by 180 Myr. Our simple test particle simulation shows that the disk in certain azimuthal regions could still keep the memory of the phase space signature from the earlier perturbation, eventually leading to the formation of the apparent two-armed phase spiral.

To better understand this process, we focus on the particles within $1500<J_{\phi}<1600\, \text{km}\, \text{s}^{-1}\, \text{kpc}$ and azimuthal ranges of $85^{\circ}<\theta_\phi<105^{\circ}$ and $125^{\circ}<\theta_\phi<145^{\circ}$ (as shown in the first and third panels of Fig.\,\ref{fig:sim_re}) to trace their trajectories back to earlier times. The right panel of Fig.\,\ref{fig:6} shows the face-on view at 200 Myr after the second perturbation, with green and blue dots representing the particles in different azimuthal ranges. The left and middle panels display the face-on views of the simulation at the times of the first and second perturbations, respectively. Fig.\,\ref{fig:7} shows the physical distance distribution of the green and blue particles from the impact site of the corresponding perturbation. At the first stage, the green and blue particles form regular phase spirals after the first perturbation. However, during the second perturbation, the majority of the green particles are much closer to the satellite's impact site ($\sim 3$ kpc), experiencing a stronger perturbation than the blue particles. As a result, the blue particles mainly preserve the single-arm phase spiral generated from the first impact as shown in the third panel of Fig.\,\ref{fig:sim_re}, while the green particles display the apparent two-armed phase spiral imprinted by both impacts in the left panel of Fig.\,\ref{fig:sim_re}.

We present the evolution of the phase spiral in different azimuthal ranges with $1500<J_\phi <1600\, \text{km}\, \text{s}^{-1}\, \text{kpc}$ in Fig.\,\ref{fig:8}, where panels with the apparent two-armed phase spiral (or secondary branch) are highlighted with blue boxes. As time progresses from the top to bottom rows, the apparent two-armed phase spiral propagates with $\theta_\phi$ as stars rotate in the disk. However, the phase mixing does not support the long-term persistence of the apparent two-armed phase spiral. Instead, it gradually transitions into a single-arm phase spiral with a weaker secondary branch after about 100 Myr, as shown in the bottom row of Fig.\,\ref{fig:8} for T=740 Myr.

%% file: discussion.tex
\subsection{Comparison with Previous Works}

\citet{Hunt2022} used two different N-body simulations (an isolated disk and an interaction one) to explore the formation of the two-armed phase spiral in the inner disk, demonstrating that the bar plays a significant role. Recently, \citet{LiC2023} suggested that the decelerating bar could give rise to the intrinsic two-armed phase spiral; the intrinsic two-armed phase spiral first emerges in the inner disk, which gradually migrates outward as the bar grows and slows down. Moreover, \citet{Asano2025} and \citet{Chiba2025} both successfully reproduced the intrinsic two-armed phase spiral through the breathing mode, using $N$-body simulations and test particle simulations, respectively. We also employ a method similar to \citet{Gandhi2022}, who used the look-back orbit to demonstrate that a single perturbation can generate distinct phase spirals in the radial direction, which are excited at different times during the impact event.

As shown in previous intrinsic two-armed phase spiral work \citep{Hunt2022, LiC2023, Chiba2025}, the intrinsic two-armed phase spiral generated by the bar seems quite symmetric, with the phase angle of the two branches separated by $\pi$. It is naturally expected that the perturbation times derived from each individual branch should be similar. However, in the observational data, as shown in Section 3, the two-armed phase spirals may not be that symmetric and may show azimuthal variation, which is an apparent two-armed phase spiral. \citet{Alinder2024} also found asymmetry between the two arms of the phase spiral in the observational data when analyzing its rotation pattern in the azimuthal region. In addition, the perturbation times that we derived from the two branches of the phase spiral are approximately 500 Myr and 320 Myr, respectively, with a difference of about 180 Myr. The perturbation times may not be accurate because of the ignorance of the self-gravity in the measurement. However, the systematic differences between the perturbation times derived from different branches should still persist. This appears to be inconsistent with the predictions of the decelerating bar model \citep{LiC2023} and the spiral arms with diffusion model \citep{Chiba2025}, both of which generate a breathing mode that eventually leads to a symmetric two-armed phase spiral. Nevertheless, in the observation, some symmetric two-armed phase spirals are still observed in the lower angular momentum (e.g., panels g in Fig.\,5 in \citet{Alinder2024} with $1400< J_\phi <1500\, \text{km}\, \text{s}^{-1}\, \text{kpc}$). This suggests that both the overlapping bending waves scenario (suggested in this work) and the breathing mode scenario induced by the bar may coexist in the Milky Way disk, given its complex perturbation history.

\citet{Asano2025} revealed a mechanism for producing two-armed phase spirals via $N$-body simulations, where a satellite interaction induces a spiral arm, triggers a breathing mode, and leads to the formation of a two-armed phase spiral in the $Z$–$V_Z$ phase space. The resulting structure in their simulations resembles our findings, showing similar asymmetry and the emergence of a secondary branch during its evolution. However, our test-particle simulations, which neglect the self-gravity, do not allow us to fully capture such self-gravity dynamical modes. Instead, the vertical structures we observe should be interpreted as coherent kinematic responses, arising from the phase wrapping after multiple external perturbations, rather than the breathing modes in a self-gravitating disk.

\subsection{Implication on the Milky Way Assembly History}

 In this work, we try to employ two consecutive external perturbations to generate the apparent two-armed phase spiral. The Milky Way may have experienced not only the Sgr dwarf galaxy encounter, which has been believed to have created the phase spiral, but also other recent and ongoing perturbations \citep[e.g.][]{Donlon2024}, interactions with dark matter subhalos \citep{Tremaine2023, Daniel2024}, or the dark matter halo wake \citep{Grand2023}. In this scenario, it seems that the more recent perturbation could not completely erase the phase spiral signature left by the earlier perturbation event. Recently, \citet{Rashid2025} also shows that the phase spiral in the inner disk is less wrapped compared with the simulation in the angular momentum space, suggesting a more recent perturbation in the inner disk. 

According to our test particle simulation, the second perturbation should occur within a relatively short time ($\sim$180 Myr) after the first perturbation. Additionally, the second perturbation should be a recent one, as the two-armed phase spiral cannot persist for a long period, as shown in Fig.\,\ref{fig:8}. Note that our results are derived from the test particle simulation without considering the self-gravity, which has been suggested as an important factor in the phase spiral evolution and its duration \citep{Darling2019}. In $N$-body simulations, the phase mixing process could occur at a much slower rate compared to the test-particle simulation due to the influence of self-gravity \citep{Widrow2023}. The perturbation time extracted from the $\theta_z-\Omega_z$ space could be underestimated.
In addition, earlier N-body studies such as \citet{DOnghia2016} and \citet{Hanies2019} have shown that satellite interactions can excite vertically coherent oscillations with azimuthal structure that last several hundred Myr.
In the future, with larger radial and azimuthal coverage of the Galactic disk observation, the phase spiral features can be much better revealed by combining the observation with high resolution N-body simulations. By then, we can better constrain the phase spiral evolution and the Milky Way perturbation history.

%% file: conclusion.tex
In this paper, we present Gaia DR3 RVS data segmented into $J_{\phi}$ and $\theta_{\phi}$ directions. We observe distinct $Z-V_Z$ phase spiral features along the $\theta_{\phi}$ direction, and the combination of these features gives rise to the formation of either an apparent two-armed phase spiral or the secondary branch spirals. Through the estimation of perturbation time, we identify a time difference between the different phase spiral features with a variation in the azimuth. Consequently, our measurements of the apparent two-armed phase spiral and the perturbation time estimations suggest that the emergence of the apparent two-armed phase spiral originates from the azimuthal overlapping of two one-armed phase spiral patterns induced by two perturbations at different times ($\sim 180$ Myr in the estimation). Self-gravity is neglected in the measurement, leading to an underestimation of the perturbation time. However, the systematic differences in the time estimation between different phase spiral branches should still persist.

To test this scenario, we conduct a test-particle simulation involving two impulsive satellite passages at different locations in the disk. The simulation shows that the second perturbation strongly affects particles in some azimuthal regions, while its influence is weaker in others, allowing the phase spiral from the first perturbation to persist. The resulting structures closely resemble those observed in Gaia DR3, where the apparent two-armed phase spiral arises from the overlap of vertical responses in different regions. We also set the time interval between the two perturbations in the simulation to be approximately $180$ Myr, and derive a similar separation ($\sim150$ Myr) between the phase spirals across different azimuthal sectors.

It is important to emphasize that the phase spirals in our test-particle simulations are kinematic responses in nature, as the self-gravity is not included. Consequently, the vertical structures we observe should not be interpreted as true breathing modes. The absence of self-gravity also implies that the evolution and persistence of these features may differ from those in full $N$-body simulations, where self-gravitating disks can sustain long-lived bending and breathing waves. Our test-particle approach likely underestimates the mixing timescales and damping behavior. Future high resolution $N$-body simulations will be essential to reveal its formation and evolution.

As the possible perturbation sources, the Sgr satellite combined with another bending wave perturbation could help to generate the apparent two-armed phase spiral. such as a more recent satellite impact, interactions with dark matter subhalos, or the influence of the dark matter halo wake, could also contribute to its formation. In the future, with a larger sample in the inner disk region, we can estimate the perturbation time for these data to check if the perturbation times have a systematic bias varied with the azimuthal direction. Moreover, we may expect a new perturbation source to affect mainly the inner disk. However, intrinsic two-armed phase spirals can still be observed in the lower angular momentum space, possibly induced by the bar. This suggests that the entire disk is not dominated by a single mechanism but rather by multiple mechanisms. A broader range of data (e.g., we look forward to the fourth Gaia data release, which will provide new RV data for the fainter stars) is still needed in the future to verify all these different mechanisms.
